УДК 004.9:66.013.512

# ЭВОЛЮЦИЯ ПАРАМЕТРИЧЕСКИХ МОДЕЛЕЙ ЧЕРТЕЖА (МОДУЛЕЙ) В САПР РЕКОНСТРУКЦИИ ПРЕДПРИЯТИЙ

В.В. Мигунов[1]

Обсуждается развитие методов автоматизации подготовки чертежей на основе так называемых модулей, содержащих одновременно параметрическое представление части чертежа и соответствующие геометрические элементы. Как этапы эволюции модульной технологии автоматизации проектирования рассматриваются варианты применения модулей для простого объединения элементов без параметрического представления с возможностью комментирования, для генерации условных графических обозначений в схемах автоматизации и чертежей трубопроводов, для хранения специфицирующих свойств изделий, для разработки специализированных частей проекта: аксонометрические схемы, профили наружных сетей и др.

## Введение

В традиционном понимании системы автоматизированного проектирования (САПР) берут на себя решение сложных и трудоемких расчетных, компоновочных и других специальных задач, и служат, в частности, целям перехода к безбумажным технологиям. Этими свойствами в полной мере обладают современные САПР, применяемые в электронной, аэрокосмической, автомобильной промышленности, в других отраслях машиностроения и др. Говорить на этом фоне об автоматизации подготовки чертежей, казалось бы, не актуально. Однако практика показывает, что с развитием предприятий неразрывно связаны и имеют большое практическое значение задачи их реконструкции, где применение различных методов математического моделирования сильно ограничено, а реальным источником сведений для производства строительно-монтажных работ являются бумажные чертежи.

Как отмечалось, например, в [Пучкова и др., 1993] в связи с пятилетним планом развития хлебопекарной промышленности, "Новые мощности предполагается вводить в основном в результате расширения, реконструкции и технического перевооружения действующих предприятий (75-80% вводимых мощностей). Доля нового строительства составит 20-25%". Эта ситуация не изменилась и к 2005 году [Ревзин, 2005]: "Немалая часть работ, выполняемых организацией, приходится на проекты реконструкции или небольшие заказы, где применение технологий трехмерного проектирования не оправдано ни по затратам, ни по срокам. Кроме того, не следует забывать, что конечный продукт работы проектировщика - это прежде всего чертежи и спецификации, а не трехмерные модели. Так что и после внедрения трехмерных технологий основная часть работ выполняется в двумерном формате". При этом известно [Гришин и др.,

[1] *420088, Казань, ул. Губкина, 50, ЦЭСИ РТ при КМ РТ, vmigunov@csp.kazan.ru*

2000, Иноземцев и др., 2000], что даже в машиностроении "Средства автоматизации конструкторского труда ... обеспечивают совершенно разную степень автоматизации в случае создания нового проекта и в случае изменения существующего". Не случайно, подводя итог впечатлениям от работы с моделирующей системой PLANT-4D, автор [Трубицын, 2004] напоминает: "Впрочем, основной задачей проектирования является не создание эффектной трехмерной картинки, а формирование полного комплекта чертежей, необходимых для строительства". В рыночных условиях в России эта тенденция выразилась в том, что "...количество проданных лицензий КОМПАС-График в три раза выше, чем КОМПАС-3D ... Многим заказчикам, в силу их производственных задач, 3D-моделирование пока просто не нужно" [АСКОН, 2005] - и это для систем, изначально ориентированных на потребности машиностроения, где трехмерное моделирование применяется широко и эффективно.

Таким образом, задача автоматизации подготовки бумажных чертежей остается актуальной. На путях ее решения стоят жесткие ограничения по уровню применимости различных моделей, проистекающие из требования минимизации трудозатрат проектировщика. У него нет времени создавать трехмерную модель, а иногда это и бессмысленно - для технологических схем, схем автоматизации технологических процессов, принципиальных электрических схем, входящих в комплект основных рабочих чертежей системы проектной документации для строительства (СПДС). Соответственно различные чертежи, а также и различные их части имеет смысл моделировать для целей автоматизации работ на различном уровне.

В данной работе показывается, как на основе единой методологии параметрических представлений частей различных чертежей в виде так называемых модулей удается автоматизировать ряд работ проектировщика, выполняемых при подготовке проектов реконструкции предприятий.

## Представление части чертежа в виде модуля

Для САПР, направленных на создание чертежей, автоматизация работ на основе специального моделирования части чертежа оправдана в случаях, когда трудоемкость ввода параметров модели существенно ниже трудоемкости непосредственного черчения с помощью графического ядра САПР, включая проведение необходимых расчетов. Например, когда требуются трудоемкие расчеты или когда нормативные требования к чертежу порождают большое количество графических изображений по малому количеству исходных данных. В обоих случаях наиболее предпочтительна параметрическая генерация чертежа по исходным данным. Для решения задач такого класса разработана технология, основанная на совместном хранении в одном элементе чертежа, называемом "Модуль", как исходных данных, так и результатов параметрической генерации.

Модуль включает видимую в чертеже совокупность геометрических элементов и невидимое в чертеже параметрическое представление моделируемого объекта. Процедуры работы с модулем помещаются в программный код САПР.

Геометрическая часть перегенерируется при всяком изменении параметров - параметрическое представление первично и полностью определяет геометрию. Структурирование параметров в модуле задается признаком его типа. Каждому типу модулей соответствует допустимое множество свойств.

Параметрические представления объектов в модулях вместе со специализированным программным кодом САПР позволяют реализовать самые разные модели объектов и методов их разработки. Специализация модулей легко опознается в чертеже, и возникает возможность многоэтапной разработки объектов проектирования и использования объектов - прототипов. Высокая структурированность параметров обеспечивает быстрый доступ к ним для различной обработки. Например, в модуль аксонометрической схемы можно автоматически вставить оси строительной подосновы, лишь выбрав модуль подосновы в чертеже. Легко автоматизируется сбор сведений при генерации спецификаций, при контроле дублирования позиционных обозначений.

Комплект параметров модулей ряда объектов может записываться на диск (без геометрических элементов), порождая информационную среду проектирования в виде библиотек прототипов. При выборе комплекта для чтения с диска геометрия генерируется в режиме on-line, и проектировщик легко ориентируется в прототипах.

Как элемент чертежа модуль подчиняется обычным правилам: его можно удалить, подвинуть, растянуть и т.д., к его геометрическим элементам возможны привязки, он может быть помещен в графическую библиотеку. Внутри модуля для входящих в него геометрических элементов хранятся габариты по зонам чертежа, позволяющие игнорировать при выводе на экран ненужные элементы. Все элементы модуля лежат на одном слое, но могут иметь разные типы линий и цвет.

При задании параметрического представления используется как стандартный пользовательский интерфейс (меню, формы ввода...), так и специализированный для черчения и корректировки.

## Варианты автоматизации работ с помощью модулей

Ниже приводятся примеры автоматизации работ по подготовке чертежей, реализуемой на основе модулей различных типов. У части из описываемых модулей указаны также допустимые свойства в квадратных скобках.

Модуль "Пользовательский" ["Привязка", "Симметрия", "Комментарий"] позволяет объединить группу элементов чертежа в один с тем, чтобы удобнее было его переносить, поворачивать, помещать в библиотеку, привязывать к другим элементам чертежа по заранее заданным осям, хранить примечания по необходимым доработкам.

Модуль "Трубопровод" автоматизирует вычерчивание трубопроводов заданного диаметра с гнутыми и сварными изгибами, с последующим нанесением обозначений арматуры, опор и поперечных сечений.

Модуль "Арматура" ["Привязка", "Симметрия", "Комментарий", "Строительная длина", "Обозначение", "Наименование", "Масса", "Примечание", "Dy", "Py"] обеспечивает удобство нанесения обозначений трубопроводной арматуры на трубопроводы с последующей генерацией специфицирующих документов.

Модуль "Таблица КИПиА" автоматизирует генерацию таблицы расположения средств автоматизации по таким параметрам, как длина граф, высота и текст каждой графы.

Модуль "Прибор" ["Привязка", "Несущая геометрия", "Позиционное обозначение", "Обозначение", "Наименование", "Масса", "Примечание", "Тип, марка оборудования", "Единица измерения", "Код единиц измерения", "Код завода-изготовителя", "Код оборудования, материала", "Цена", "Наименование и технич. х-ка", "На щите", "Функциональный признак прибора", "Верхний индекс", "Нижний индекс", "Комментарий", "Тип линии приборов КИП"] и аналогичный ему модуль "Исполнительный механизм" служат целям генерации условных графических обозначений по требованиям [ГОСТ, 1985], автоматизации их специфицирования, а также нанесения линий связи с привязками по требованиям [РМ, 1992]. Обеспечивается прием данных из электронных каталогов выпускаемых изделий. После выбора в каталогах новые данные используются для перегенерации соответствующей части чертежа.

Модуль "План этажа" ["Комментарий", "Параметры этажа в плане", "Масштаб при создании"] автоматизирует большинство операций по вычерчиванию планов строительной подосновы: этажей, фундаментов, покрытий и перекрытий, а также служит основой для построения разрезов строительной подосновы. Подоснова включает координационные оси зданий и сооружений, колонны, перегородки, проемы и другие строительные конструкции, входящие в состав чертежей различных марок СПДС согласно [ГОСТ, 1997].

Модуль "Обозначение для аксонометрии" ["Привязка", "Симметрия", "Комментарий", "Вырезаемый на трубе отрезок"] позволяет автоматически наносить на трубы в аксонометрических схемах условные графические обозначения арматуры и элементов трубопроводов по требованиям с учетом требования, что через их плоскость требуется провести выносную линию размера вдоль одной из осей координат.

Модуль "Аксонометрическая схема" ["Комментарий", "Параметры аксонометрич. схемы", "Масштаб при создании"] автоматизирует большинство операций по созданию аксонометрических схем трубопроводных систем в соответствии с требованиями [ГОСТ, 1988, ГОСТ, 1983, ГОСТ, 1981].

Модуль "Оформление чертежа" ["Описание оформления чертежа"] служит для вычерчивания рамок, основной и дополнительных надписей и позволяет легко изменить формат чертежа, например с А1 на А0, с соблюдением всех требований ГОСТов, а также стандартов предприятия.

"Табличный" модуль ["Комментарий", "Описание таблицы"] автоматизирует все операции создания специфицирующих таблиц, включая их автоматическое заполнение из модулей типа "Позиционное обозначение", "Профиль наружной сети ВК", "Аксонометрическая схема", настройку параметров граф, строк, текстов, расположения продолжений, прием данных из электронных каталогов изделий, выпускаемых промышленностью, а также специальные задачи специфицирования узлов (например, фланцевых соединений).

Модуль "Позиционное обозначение" ["Тип позиционного обозначения", "Тип объекта позиционного обозначения", "Специфицирующие свойства"] служит целям автоматизации специфицирования чертежей, храня в каждом позиционном обозначении специфицирующую информацию для последующей генерации спецификаций; обеспечивает прием данных из электронных каталогов изделий, выпускаемых промышленностью.

Модуль "Профиль наружной сети ВК" ["Комментарий", "Параметры профиля наружной сети ВК", "Масштаб при создании"] автоматизирует большинство операций по вычерчиванию профилей наружных сетей водоснабжения и канализации по требованиям [ГОСТ, 1982].

Модуль "Молниезащита зданий и сооружений" ["Комментарий", "Параметры молниезащиты зданий и сооружений", "Масштаб при создании"] автоматизирует проектирование молниезащиты зданий и сооружений, в том числе on-line расчет сечений зон защиты по требованиям [РД, 1987, СО, 2004] при изменении положения или высоты молниеприемников движением курсора мыши.

Модуль "Электронная подпись" ["Сотрудник", "Должность", "Пароль", "Дата", "Время"] решает задачу контроля целостности (неизменности с момента подписания) чертежа.

Имеются также рабочие модули, которые помещаются в чертеж только на время модификации параметрических представлений строительной подосновы, аксонометрических схем, профилей наружных сетей, проектов молниезащиты. Эти модули служат целям ускорения работы. Они содержат, кроме геометрической части, сведения о типе объекта и его номере во внутреннем списке таких объектов в параметрическом представлении. Например, элемент номер 3 списка "Размеры радиусов горизонтальных сечений" в проекте молниезащиты. Если выполняется операция изменения размеров радиусов сечений, то проектировщик автоматически сможет выбрать в чертеже только требуемые размеры.

Наиболее развитое применение модулей осуществляется в специализированных расширениях САПР, реализованных на основе модулей "План этажа", "Аксонометрическая схема", "Табличный", "Профиль наружной сети ВК", "Молниезащита зданий и сооружений" в рамках модульной технологии разработки расширений САПР [Мигунов, 2004]. Во всех этих случаях работа по модификации параметрических представлений ведется в отдельном основном меню, а в чертеж помещаются временные рабочие модули.

## Заключение. Развитие параметрических моделей чертежа

Охарактеризованные выше параметрические модели частей чертежа реализованы в процессе развития САПР реконструкции предприятий TechnoCAD GlassX, последняя версия которой называется TechnoCAD Glass [Мигунов, 2006]. При этом автоматизация работ проектировщика вначале сводилась к генерации изображений по параметрам, задаваемым проектировщиком в формах ввода и путем выбора в меню возможных вариантов. С появлением электронных каталогов изделий, выпускаемых промышленностью, появилась возможность назначать параметры путем выбора в этих каталогах [Мигунов, 2005].

Состав свойств, облегчающих работу, наращивался постепенно. Свойство "Привязка" позволило устанавливать нужные модули точно в нужное место и с нужной ориентацией. Свойство "Симметрия" уменьшило число возможных вариантов ориентации обозначений арматуры в аксонометрических схемах. Хранение в модулях специфицирующих свойств изделий позволило автоматизировать заполнение табличных конструкторских документов путем сбора сведений из различных модулей, находящихся в текущем чертеже либо во множестве чертежей на диске, выбираемых проектировщиком [Мигунов, 2005].

Наиболее комплексная автоматизация работ достигается при использовании параметрических моделей значительной части чертежа, объединяющей сильно связанные по смыслу элементы. Здесь удается автоматизировать практически все операции вычерчивания. Параметрическое представление в этом случае содержит списки объектов со ссылками принадлежности, образующие реляционную базу данных, а также установки вычерчивания элементов [Мигунов и др., 2003a, Мигунов и др., 2003b]. Вместе с программным кодом соответствующего специализированного расширения САПР эти параметрические представления образуют аналог объекта в терминологии объектно-ориентированного программирования. По мере изменения СПДС и других нормативных требований разработанные в рамках модульной технологии параметрические модели позволяют легко учитывать новые требования [Мигунов, 2006].

## Список литературы

**[АСКОН, 2005]** АСКОН: итоги 2004 года, стратегия 2005-2007//САПР и графика, 2005, специальный выпуск "Аскон".

**[ГОСТ, 1981]** ГОСТ 21.602-79 (1981) СПДС. Отопление, вентиляция и кондиционирование воздуха. Рабочие чертежи. – М.: Госстандарт, 1981.

**[ГОСТ, 1982]** ГОСТ 21.604-82 СПДС. Водоснабжение и канализация. Наружные сети. Рабочие чертежи. – М.: ГОССТАНДАРТ, 1982.

**[ГОСТ, 1983]** ГОСТ 21.601-79 (1983) СПДС. Водопровод и канализация. Рабочие чертежи. – М.: Госстандарт, 1983.

**[ГОСТ, 1985]** ГОСТ 21.404-85 СПДС. Автоматизация технологических процессов. Обозначения условные приборов и средств автоматизации в схемах. – М.: Госстандарт, 1985.

**[ГОСТ, 1988]** ГОСТ 21.401-88 СПДС. Технология производства. Основные требования к рабочим чертежам. – М.: Госстандарт, 1988.

**[ГОСТ, 1997]** ГОСТ 21.101-97 СПДС. Основные требования к проектной и рабочей документации. – М.: МНТКС, 1997.

**[Гришин и др., 2000]** Гришин С.А., Долгов Д.В. Выбор САПР средств технологического оснащения//Автоматизация и информатизация в машиностроении (АИМ 2000). Сборник трудов Первой международной электронной научно-технической конференции. – Тула: ТулГУ, 2000.

**[Иноземцев и др., 2000]** Иноземцев А.Н., Троицкий Д.И. Автоматическая параметризация машиностроительных чертежей//Автоматизация и управление в машиностроении, № 15, 2000 [Электронный журнал]. (с) СЦ НИТ, Designed by Wild Rose Studio. – М.: МГТУ "Станкин". – Режим доступа: http://magazine.stankin.ru, свободный. – Загл. с экрана. – Яз. рус.

**[Мигунов и др., 2003a]** Мигунов В.В., Кафиятуллов Р.Р., Сафин И.Т. Модульная технология разработки расширений САПР: молниезащита зданий и сооружений//Известия Тульского государственного университета. Серия "Математика. Механика. Информатика". – Т.9. Вып. 3. Информатика. – Тула: Изд-во ТулГУ, 2003.

**[Мигунов и др., 2003b]** Мигунов В.В., Кафиятуллов Р.Р., Сафин И.Т. Модульная технология разработки расширений САПР: профили наружных сетей водоснабжения и канализации//Известия Тульского государственного университета. Серия "Математика. Механика. Информатика". – Т.9. Вып. 3. Информатика. – Тула: Изд-во ТулГУ, 2003.

**[Мигунов, 2004]** Мигунов В.В. Модульная технология разработки проблемно-ориентированных расширений САПР реконструкции предприятия//Известия Тульского государственного университета. Серия "Экономика. Управление. Стандартизация. Качество". – Вып. 1. – Тула: Изд-во ТулГУ, 2004.

**[Мигунов, 2005]** Мигунов В.В. Специфицирование в комплексной САПР реконструкции предприятий на основе модулей в чертеже и электронных каталогов //Информационные технологии в проектировании и производстве, 2005, № 1.

**[Мигунов, 2006]** Мигунов В.В. Развитие комплексной САПР реконструкции предприятий TechnoCAD Glass: молниезащита и защита информации//САПР и графика, 2006, № 1.

**[РД, 1987]** РД 34.21.122-87. Инструкция по устройству молниезащиты зданий и сооружений. – М.: Министерство энергетики и электрификации СССР, 1988.

[Ревзин 2005] Ревзин В.Е. Комплексная автоматизация проектных организаций: цели, условия, результаты//CADmaster, 2005, № 4. – М.: Consistent Software.

**[РМ, 1992]** РМ4-2-92. Системы автоматизации технологических процессов. Схемы автоматизации. Указания по выполнению. – М.: ГПКИ Проектмонтажавтоматика", 1992.

**[Трубицын, 2004]** Сергей Трубицын. Сложность и комплексы или простота и комплексность?//CADmaster, 2004, № 3. – М.: Consistent Software.

**[Пучкова и др., 1993]** Пучкова Л.И., Гишин А.С., Шаргородский И.И., Черных В.Я. Проектирование хлебопекарных предприятий с основами САПР. – М.: Колос, 1993.

**[СО, 2004]** СО-153-34.21.122-2003. Инструкция по устройству молниезащиты зданий, сооружений и промышленных коммуникаций. – М.: Издательство МЭИ, 2004.